\documentclass[prb,twocolumn,amsmath,amssymb,superscriptaddress]{revtex4}

\usepackage{graphicx}

\usepackage{epstopdf} \DeclareGraphicsExtensions{.eps, .pdf}

\usepackage{epsfig,psfrag,amsmath,amssymb,float}
\begin{document}

\newcommand{\ltwid}{\mathrel{\raise.3ex\hbox{$<$\kern-.75em\lower1ex\hbox{$\sim$}}}} \newcommand{\gtwid}{\mathrel{\raise.3ex\hbox{$>$\kern-.75em\lower1ex\hbox{$\sim$}}}} \def\K{{\bf{K}}} \def\Q{{\bf{Q}}} \def\Gbar{\bar{G}} \def\tk{\tilde{\bf{k}}} \def\k{{\bf{k}}} \def\kt{{\tilde{\bf{k}}}} \def\p{{\bf{p}}} \def\q{{\bf{q}}} \def\pp{{\bf{p}}^\prime} \def\Gpp{\Gamma^{pp}} \def\Phid{\Phi_d(\k,\omega_n)} \def\ld{\lambda_d(T)} \def\n{\langle n \rangle } \def\dw{d_{x^2-y^2}} \def\Ub{\bar{U}}

\title{Spin Susceptibility Representation of the Pairing Interaction for the   two-dimensional Hubbard Model}

\author{T.A.~Maier} \affiliation{Computer Science and Mathematics Division,\\ Oak Ridge   National Laboratory, Oak Ridge, TN 37831-6164} \email{maierta@ornl.gov}

\author{M.~Jarrell} \affiliation{Department of Physics,\\ University of Cincinnati,   Cincinnati, OH 45221} \email{jarrell@physics.uc.edu}

\author{D.J.~Scalapino} \affiliation{Department of Physics,\\ University of California,   Santa Barbara, CA 93106-9530} \email{djs@vulcan2.physics.ucsb.edu}

\date{\today}

\begin{abstract}
  Using numerical dynamic cluster quantum Monte Carlo results, we study a simple   approximation for the pairing interaction of a two-dimensional Hubbard model with an   on-site Coulomb interaction $U$ equal to the bandwidth. We find that with an effective   temperature dependent coupling $\Ub(T)$ and the numerically calculated spin   susceptibility $\chi(K-K')$, the d-wave pairing interaction is well approximated by   $\frac{3}{2} \Ub^2\chi(K-K')$.

\end{abstract}

\maketitle

% ==========BODY OF PAPER =========================================

\section{Introduction}

Numerical calculations have shown that the dominant contribution to the d-wave pairing interaction in the Hubbard model comes from a spin $S=1$ channel \cite{maier:pairmech,maier:pairint}. It is therefore interesting to determine how well the simple RPA form
\begin{equation}
  \label{eq:1}
  \frac{3}{2}\Ub^2\chi(K-K')
\end{equation}
can describe this interaction. Here $\chi$ is the spin susceptibility, $\Ub$ is an effective coupling strength and $K=(\K,\omega_n)$ with $\omega_n=(2n+1)\pi T$ a fermion Matsubara frequency. Eq.~\eqref{eq:1} is the form that one expects in weak coupling \cite{berk:66,scalapino:95,Ueda:00,chubokov:03}. However, we are interested in the case in which the Hubbard on-site Coulomb interaction $U$ is comparable to the bandwidth, since this is the region which is believed to be appropriate for the cuprates \cite{tremblay:06}. This is also the parameter regime in which the pairing is the strongest \cite{maier:pairint}. Here, for 
the parameters we use throughout this paper including a near neighbor hopping 
$t = 1$ and $U = 8$, the single-particle propagator and density of states show 
clear evidence that one is dealing with a doped Mott system.  In particular one sees structures associated with upper and lower Hubbard bands as well as a conduction band \cite{preuss:95,maier:pairint}. In this case, it is unclear whether the simple form given by Eq.~\eqref{eq:1} can adequately describe the pairing interaction. We believe that there are two reasons that this is possible. First, the leading contribution of $U$ to the d-wave pairing channel vanishes due to the pairing symmetry and secondly, the dressed spin susceptibility $\chi$ and not the perturbative RPA form will be used in Eq.~\eqref{eq:1}.

To see how well Eq.~\eqref{eq:1} can describe the pairing interaction, we first discuss the fitting procedure which will be used to determine a temperature dependent effective coupling $\Ub(T)$. Following this we compare the d-wave eigenvalue and the $\K$ and $\omega_n$ dependence of the eigenfunction of the Bethe-Salpeter equation using the interaction given by Eq.~\eqref{eq:1} with the results obtained using the "exact" interaction. Throughout this paper, "exact" will refer to numerical results obtained using a dynamic cluster approximation (DCA) Monte Carlo technique \cite{hettler:dca1,hettler:dca2,jarrell:dca3,maier:rev}. We will then conclude by discussing the relationship of $\Ub$ to the bare interaction.

\section{Fitting the Effective Coupling \boldmath$\Ub$}

For the traditional low $T_c$ superconductors, the phonon mediated s-wave pairing interaction is characterized by \cite{scalapino:parks69}
\begin{equation}
  \label{eq:2}
  \alpha^2 F(\omega) = -\sum_\nu \frac{\langle |g^2_{\p,\p'}|
    \frac{\Im m}{\pi}
    D_\nu(\p-\p',\omega)\delta(\varepsilon_\p)\delta(\varepsilon_{\p'})\rangle_{\p\p'}}{\langle
    \delta(\varepsilon_\p)\rangle_\p}
\end{equation}
Here, $D_\nu(\q,\omega)$ is the phonon propagator for polarization $\nu$ and $\langle \delta(\varepsilon_\p)\rangle_\p$ represents an average over the Fermi surface weighted by a $\p$-dependent density of states which varies as $v^{-1}_F(\p)$. The coupling strength $\lambda$ is given by
\begin{equation}
  \label{eq:3}
  \lambda=2\int^\infty_0 d\omega \frac{\alpha^2F(\omega)}{\omega}
\end{equation}
and it clearly depends upon the phonon dynamics. However, by substituting Eq.~\eqref{eq:2} into Eq.~\eqref{eq:3} and making use of the usual dispersion relation, $\lambda$ can be conveniently expressed in terms of the zero frequency limit of the phonon propagator $D_\nu(\q,\omega=0)$
\begin{equation}
  \label{eq:4}
  \lambda=-\sum_\nu \frac{\langle |g^2_{\p,\p'}|
    D_\nu(\p-\p',0)\delta(\varepsilon_\p)\delta(\varepsilon_{\p'})\rangle_{\p\p'}}{\langle
    \delta(\varepsilon_\p)\rangle_\p}\,.
\end{equation}
At a finite temperature $D_\nu(\p-\p',0)$ is replaced by $1/2[D_\nu(\p-\p',0)+D_\nu(\p-\p',2\pi T)]$.

For the Hubbard model we lack a Migdal theorem and the pairing interaction is calculated with a dynamic cluster quantum Monte Carlo algorithm \cite{hettler:dca1,hettler:dca2,jarrell:dca3,maier:rev}. The dynamical cluster approximation maps the original lattice model onto a periodic cluster of size $N_c$ sites embedded in a self-consistent host. The essential assumption is that short-range quantities, such as the self energy and its functional derivatives (the irreducible vertex functions) are well represented as diagrams constructed from the coarse-grained Green's function. For the problem of interest, this is a reasonable assumption for systems where the correlations that mediate the pairing are short-ranged. To this end, the first Brillouin zone is divided into $N_c$ cells, with each cell represented by its center wave-vector $\K$ surrounded by $N/N_c$ lattice wavevectors labeled by $\tk$. The reduction of the $N$-site lattice problem to an effective $N_c$ site cluster problem is achieved by coarse-graining the single-particle Green's function, {\it i.e.}  averaging $G(\K+\tilde{\k})$ over the $\tk$ within a cell which converges to a cluster Green's function $G_c(\K)$. Consequently, the compact Feynman diagrams constructed from $G_c(\K)$ collapse onto those of an effective cluster problem embedded in a host which accounts for the fluctuations arising from the hopping of electrons between the cluster and the rest of the system.  The compact cluster quantities are then used to calculate the corresponding lattice quantities.

\begin{figure}[htbp]
    \centering
    \includegraphics[width=3.5in]{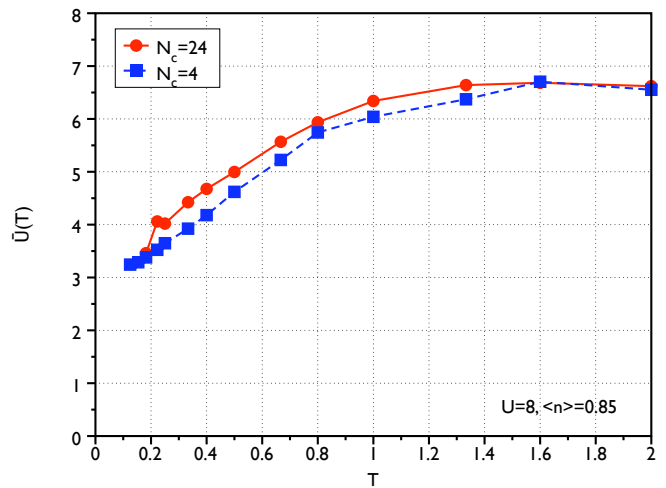}
    \caption{The coupling strength $\Ub$ for $U=8t$ and a site filling       $\n=0.85$ calculated for two cluster sizes $N_c=4$ and 24.}
    \label{fig:1}
  \end{figure}

The pairing interaction is given by the irreducible part of the particle-particle vertex
\begin{equation}
  \label{eq:5}
  \Gamma^{pp}(K;K')\equiv \Gamma^{pp}(K,-K; K',-K')
\end{equation}
with $K=(\K,\omega_n)$. One can also use the DCA to calculate the spin susceptibility $\chi(\Q,\omega_n)$ \cite{jarrell:dca3,maier:rev}. In an analogous manner to Eq.~\eqref{eq:4}, we now introduce a d-wave coupling strength
\begin{equation}
  \label{eq:6}
  -\frac{\frac{1}{2}\langle g(\K) \Gamma^{pp}_{\rm even}(\K,\pi T; \K',\pi     T)g(\K')\rangle_{\K\K'}}{\langle g^2(\K)\rangle_{\K}}
\end{equation}
with the even frequency, even momentum part of the irreducible particle-particle vertex,
\begin{eqnarray}
  \label{eq:11}
  \Gamma^{pp}_{\rm even}(\K,\pi T;\K',\pi T) &=& \frac{1}{4} \left( \right.
  \Gamma^{pp}(\K,\pi T; \K',\pi   T)\nonumber\\
  &+&\Gamma^{pp}(\K,\pi T, -\K', \pi T)\nonumber\\
  &+&\Gamma^{pp}(\K,\pi T, \K', -\pi T)\nonumber\\
  &+&\left. \Gamma^{pp}(\K,\pi T, -\K', -\pi T)\right)
\end{eqnarray}
and $g(\K)=(\cos K_x - \cos K_y)$. An effective coupling $\Ub(T)$ may then be obtained by requiring that the d-wave coupling strength in Eq.~(\ref{eq:6}) is the same at a given temperature for the approximate interaction one obtains by replacing $\Gamma^{pp}(K;K')$ by the interaction in Eq.~\eqref{eq:1}. In Fig.~\ref{fig:1} we show the results for $\Ub(T)$ for two different DCA cluster sizes\footnote{For an illustration of the 4-site   and 24-site cluster geometries, please see Fig.3 in Ref.~\cite{maier:pairint}.}, $N_c=4$ and $N_c=24$, for the case in which $U=8$ and the site filling $\n=0.85$.  Here one sees that $\Ub$ is smaller than $U$ and decreases at lower temperatures. We will discuss the physics that underlies this effect after we explore how well $\frac{3}{2}\Ub^2\chi(K-K')$ represents $\Gamma^{pp}(K;K')$.

\section{Results for the Particle-Particle Bethe-Salpeter Equation}

\begin{figure}[htbp]
  \centering
  \includegraphics[width=3.5in]{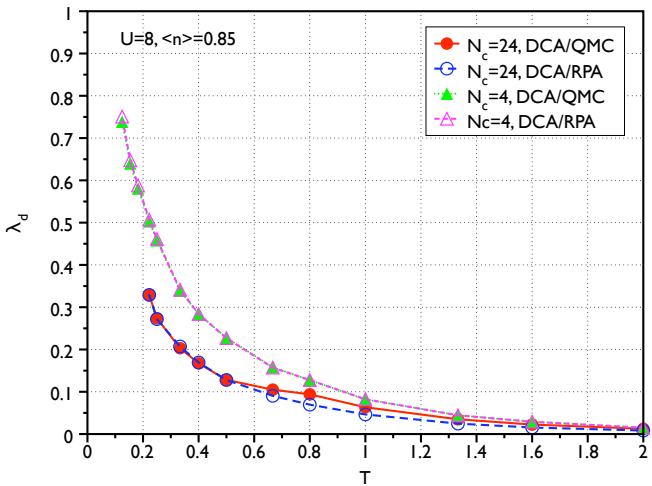}
  \caption{The d-wave eigenvalue versus $T$ obtained from the RPA form,     Eq.~(\ref{eq:1}) (open symbols) and from the "exact" DCA interaction (solid     symbols) for $N_c=4$ and 24.}
  \label{fig:2}
\end{figure}

  The leading low temperature eigenvalue of the particle-particle Bethe-Salpeter equation
\begin{eqnarray}
    -\frac{T}{N_c}\ \sum_{K^\prime} \Gamma^{\rm pp}_{\rm even} \left(K, -K; K^\prime,       -K^\prime\right)\,
    {{\bar{\chi}}_0^{\rm pp}}(K')
    \, \phi_\alpha (K^\prime) = &&\nonumber\\
    && \hspace{-2cm}\lambda_\alpha \phi_\alpha (K)
    \label{eq:7}
\end{eqnarray}
corresponds to an eigenfunction with d-wave symmetry. Here we have coarse-grained the Green's function legs, ${{\bar{\chi}}_0^{\rm pp}}(K') = \frac{N_c}{N}\sum_{\tk^{\prime}} G_\uparrow (\K^\prime+\tk^\prime) \, G_\downarrow (-\K^\prime-\tk^\prime)$, according to the DCA assumption. The curves with solid symbols in Fig.~\ref{fig:2} show the d-wave eigenvalue versus $T$ obtained from Eq.~\eqref{eq:7} with the "exact" DCA interaction $\Gamma^{pp}$. The curves with open symbols show the d-wave eigenvalue obtained from Eq.~\eqref{eq:7} when $\Gamma^{pp}$ is replaced by $\frac{3}{2}\Ub^2\chi(K-K')$. Here we are using DCA results for $\chi(K-K')$ as well as the single-particle propagator $G(\k)$ that appears in Eq.~\eqref{eq:7}. One sees that with $\Ub(T)$ determined as discussed in Sec.~2, the temperature dependence and the size of the d-wave eigenvalue are well accounted for by the simple form of the interaction given by Eq.~\eqref{eq:1}. For the 4-site cluster, the eigenvalue $\lambda_d$ is larger than that for the 24-site cluster because of the absence of pair field fluctuations for the 4-site cluster \cite{maier:schm}.

\begin{figure}[htbp]
  \centering
  \includegraphics[width=3.5in]{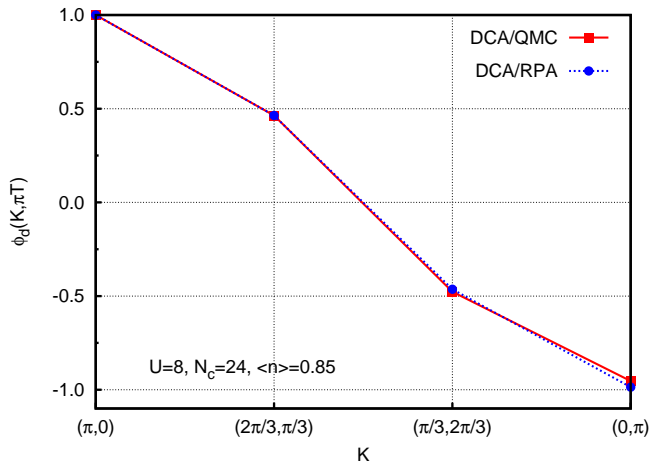}
  \caption{The momentum dependence of the d-wave eigenvector $\phi(\K,\omega_n)$ for     $\omega_n=\pi T$ and $T=0.33$ of the approximate interaction, Eq.~\eqref{eq:1},     compared to the "exact" DCA interaction $\Gamma^{pp}_{\rm even}$.     $\phi(\K,\omega_n)$ has been normalized to its value at $\K=(\pi,0)$.}
  \label{fig:4a}
\end{figure}

\begin{figure}[htbp]
  \centering
  \includegraphics[width=3.5in]{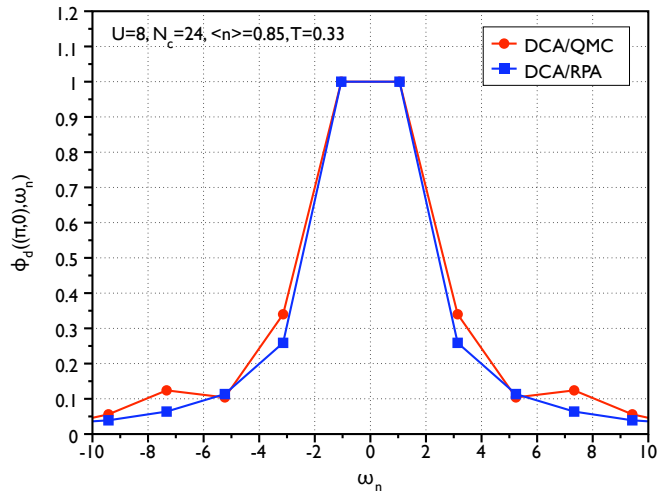}
  \caption{The frequency dependence of the d-wave eigenvector $\phi(\K,\omega_n)$     for $\K=(\pi,0)$ and $T=0.33$ of the approximate interaction, Eq.~\eqref{eq:1},     compared to the "exact" DCA interaction $\Gamma^{pp}_{\rm even}$ for two     different temperatures. $\phi(\K,\omega_n)$ has been normalized to its value at     $\omega_n=\pi T$.}
  \label{fig:4}
\end{figure}

In the following we show results for the 24-site cluster which was shown in Ref.\cite{maier:schm} to be close to convergence with respect to the temperature dependence of the pair field susceptibility. The momentum dependence of the eigenfunction obtained using the approximate form of the interaction has the same dominant $(\cos K_x - \cos K_y)$ behavior as the exact DCA result as shown in Fig.~\ref{fig:4a}.  Furthermore, as shown in Fig.~\ref{fig:4}, the Matsubara frequency dependence of the DCA and the approximate interaction are remarkably similar.

\section{The Effective \boldmath$\Ub(T)$}

By fitting $\Ub(T)$ so that the d-wave strength of the approximate interaction is equal to that of the "exact" DCA interaction, we have found that the approximate form, Eq.~\eqref{eq:1}, does an excellent job in describing the d-wave eigenvalue and eigenfunction. From a purely phenomenological point of view, this is an important result. It means that to the extent that the Hubbard model gives an appropriate description of the cuprates, inelastic neutron scattering experiments which give $\Im m \chi(q,\omega)$ provide a way of determining the momentum and frequency structure of the pairing interaction.
A similar analysis was applied to the heavy fermion superconductor 
UPT$_3$.\cite{m_norman_87,w_putikka_89}.
However, here one would like to have a better understanding of $\Ub$.

\begin{figure}[htbp]
  \centering
  \includegraphics[width=3.5in]{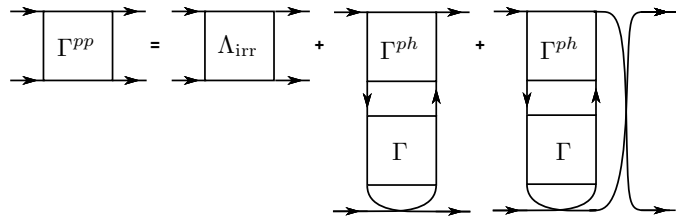}
  \caption{Decomposition of the irreducible particle-particle vertex $\Gamma^{pp}$     into a fully irreducible vertex $\Lambda_{\rm irr}$ plus reducible particle-hole     contributions in the cross channel.  Here $\Gamma$ denotes the full 2-particle     vertex and $\Gamma^{ph}$ the irreducible particle-hole vertex.}
  \label{fig:5}
\end{figure}

\begin{figure}[htbp]
  \centering
  \includegraphics[width=3.5in]{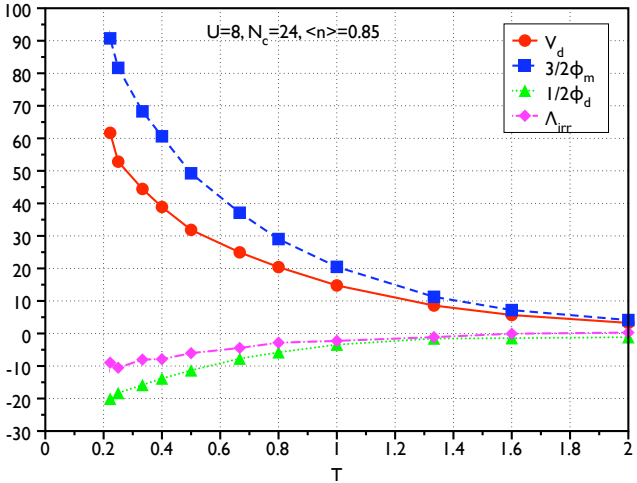}
  \caption{Decomposition of the d-wave pairing interaction $V_d$ into
    d-wave projections of the fully irreducible vertex $\Lambda_{\rm
      irr}$, and a magnetic ($S=1$) channel $3/2\phi_m$ and a charge
    ($S=0$) channel $1/2\phi_d$.}
  \label{fig:6}
\end{figure}

As previously discussed \cite{maier:pairmech,maier:pairint}, the irreducible particle-particle vertex can be decomposed into a fully irreducible two-fermion vertex $\Lambda_{\rm irr}$ plus contributions from particle-hole channels, as illustrated in Fig.~\ref{fig:5}. These particle-hole channels can be separated into density ($S=0$) and magnetic ($S=1$) channels. For the even frequency, even momentum part of $\Gamma^{pp}$ that is of interest, one has
\begin{eqnarray}
  \Gamma^{\rm pp}_{\rm even} (K,K') 
  &=&  \wedge_{\rm irr,even} (K, K')\nonumber\\
  &+&  \frac{1}{2} \Phi_d(K,K') + \frac{3}{2} \Phi_m (K,K')  
\label{eq:8}
\end{eqnarray}
with the even frequency, even momentum part of the fully irreducible two-particle vertex $\wedge_{\rm irr, even}$ and
\begin{eqnarray}
  \label{eq:9}
  \Phi_{d/m}(K,K') = \hspace*{-2.5cm}&&\\
  &&\frac{1}{2}\left[\Gamma^{red}_{d/m}(\K-\K',\omega_n-\omega_{n'};\K',\omega_{n'};-\K,-\omega_n)\right.\nonumber\\
  &+& \Gamma^{red}_{d/m}     (\K+\K',\omega_n-\omega_{n'};-\K',\omega_{n'};-\K,-\omega_n)       \nonumber\\
  &+&     \Gamma^{red}_{d/m}(\K-\K',\omega_n+\omega_{n'};\K',-\omega_{n'};-\K,-\omega_n)           \nonumber\\
  &+& \left.             \Gamma^{red}_{d/m}(\K+\K',\omega_n+\omega_{n'};-\K',-\omega_{n'};-\K,-\omega_n)\right]\,.\nonumber
\end{eqnarray}
Here, $K=(\K,\omega_n)$ and $\Gamma^{red}_{d/m}$ are the two-particle reducible contributions to the full 4-point vertex $\Gamma_{d/m}$ calculated in the S=0 (d) or S=1 (m) channel, i.e. $\Gamma^{red}_{d/m}(Q;K;K')=\Gamma_{d/m}(Q;K;K')-\Gamma^{ph}_{d/m}(Q;K;K')$ where $\Gamma^{ph}_{d/m}$ is the irreducible vertex in the corresponding channel. The center of mass and relative wave vectors and frequencies in these channels are labeled by the first, second and third arguments respectively. In Fig.~\ref{fig:6} the d-wave projection, Eq.~(\ref{eq:6}), of $\Gamma^{pp}_{\rm even}(K,K')$ along with the three contributions on the right hand side of Eq.~(\ref{eq:8}) are plotted versus the temperature. As expected, the dominant contribution comes from the magnetic ($S=1$) term. The fully irreducible channel is relatively ineffective, but at low temperatures the charge density ($S=0$) channel acts to reduce the effective d-wave pairing strength. This is one of the reasons for the decrease found in $\Ub(T)$. A similar effect is found in the simple RPA treatment where the effective pairing interaction is $V_{\rm spin}-V_{\rm charge}$. However, the RPA result for $V_{\rm charge}$ does not vary as strongly with the temperature.

\begin{figure}[htbp]
  \centering
  \includegraphics[width=3.5in]{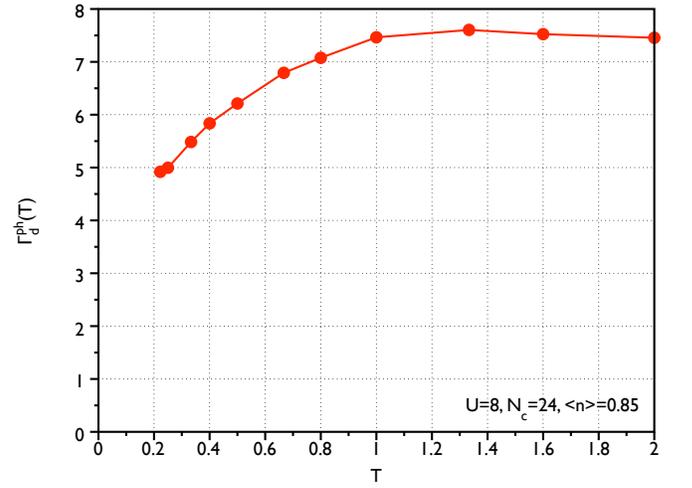}
  \caption{Coupling $\bar{\Gamma}^{ph}$ to the $S=1$ susceptibility.}
  \label{fig:7}
\end{figure}

In addition, from Fig.~\ref{fig:5}, one can see that even if one were to just consider the magnetic channel, $\Ub$ would be a more complicated object. The coupling to the $S=1$ susceptibility is given by an irreducible particle-hole vertex $\Gamma^{ph}$ which depends upon $K-K'$ as well as internal momenta and Matsubara frequencies. In Fig.~\ref{fig:7} we show results for
\begin{eqnarray}
  \label{eq:10}
  \bar{\Gamma}^{ph}=\left(\frac{\langle g(\K) \Phi_m(\K,\pi T;\K',\pi
      T) g(\K')\rangle_{\K\K'}}{\langle g(\K)
      \chi_{\rm even}(\K,\pi T;\K',\pi T) g(\K')\rangle_{\K\K'}}\right)^{1/2}\,,
\end{eqnarray}
where $\chi_{\rm even}$ is the even frequency, even momentum part of the spin susceptibility $\chi$ obtained in an analogous manner to Eq.~(\ref{eq:11}). One can see that this estimate for $\bar{\Gamma}^{ph}$ is smaller than $U$. This is consistent with the results of earlier Quantum Monte Carlo calculations \cite{bulut:93b,huang:06} which found that the electron-spin fluctuation vertex decreased with decreasing temperature for all momentum transfers.

\section{Conclusion}
\label{sec:conclusion}

To conclude, this work has shown that the momentum and frequency dependence of the d-wave pairing interaction for the Hubbard model in the parameter regime which is believed to be appropriate to the cuprates is well approximated by the spin susceptibility. The strength of the coupling when written in the RPA form $\frac{3}{2} \Ub^2(T)$ requires a temperature dependent effective $\Ub(T)$ which is reduced from the bare $U$ and decreases at lower temperature in order to phenomenologically account for the effect of the charge channel and vertex corrections.

\acknowledgments This research was enabled by computational resources of the Center for Computational Sciences at Oak Ridge National Laboratory and conducted at the Center for Nanophase Materials Sciences, which is sponsored at Oak Ridge National Laboratory by the Division of Scientific User Facilities, U.S. Department of Energy. MJ acknowledges support by NSF DMR-0312680. DJS and MJ acknowledge the Center for Nanophase Materials Science at Oak Ridge National Laboratory for support.

 \bibliography{mybib}

%\begin{thebibliography}{99}

%\end{thebibliography}

\end{document}